\documentclass[12pt]{article}
\usepackage{graphicx}
\usepackage{epsfig}
\usepackage{mathtext}
\usepackage[T2A]{fontenc}
\usepackage[koi8-r]{inputenc}
\usepackage[russian,english]{babel}
\newcommand{\beq}{\begin{equation}}
\newcommand{\eeq}{\end{equation}}
\newcommand{\beqn}{\begin{eqnarray}}
\newcommand{\eeqn}{\end{eqnarray}}
\begin{document} 
\title{\textbf{NEUTRINO - ELEMENTARY PARTICLES OR PHANTOMS}}
\author{V.P. Efrosinin\\
Institute for Nuclear Research, Russian Academy of Sciences,\\
pr. Shestidesyatiletiya Oktyabrya 7a, Moscow, 117312 Russia}

\date{}
\renewcommand {\baselinestretch} {1.3}

\maketitle

\begin{abstract}
The problem of statistical uncertainty of an estimation of parametres
 of neutrino oscillations with $\chi^2$-test is discussed. 
    
\end{abstract}

In Ref. \cite{giun} CPT theorem check has been spent to models of two-neutrino
mixing in wich short-baseline at definition of parametres of mixing
$sin^2 2\theta$ and squared-mass $\delta 
m^2_{\nu}=\mid m^2_2-m^2_1\mid$, properly:
\begin{eqnarray}
\label{eq:MM1}
P_{\nu_e \rightarrow \nu_e}=1-sin^2 2\theta_{\nu}
sin^2\Bigl(\frac{\delta m^2_{\nu}L}{4E}\Bigr), 
\end{eqnarray}
where $L$ - distance from the source to the detector.

If the CPT theorem is carried out a similar parity of (\ref{eq:MM1})
can be fair and for an $\bar{\nu}$:
\begin{eqnarray}
\label{eq:MM2}
P_{\bar{\nu}_e \rightarrow \bar{\nu}_e}=1-sin^2 2\theta_{\bar{\nu}}
sin^2\Bigl(\frac{\delta m^2_{\bar{\nu}}L}{4E}\Bigr). 
\end{eqnarray}
For calculation of parametres $\delta m^2_{\nu}$, $sin^2 2\theta_{\nu}$ and
$\delta m^2_{\bar{\nu}}$, $sin^2 2\theta_{\bar{\nu}}$ the goodness-of-fit
method was used \cite{malt}. Also asymmetries for mass and mixing are
calculated:
\begin{eqnarray}
\label{eq:MM3}
A^{CPT}_{\delta m^2}&=&\delta m^2_{\nu}-\delta m^2_{\bar{\nu}},\nonumber\\
A^{CPT}_{sin^2 2\theta}&=&sin^2 2\theta_{\nu}-sin^2 2\theta_{\bar{\nu}}.
\end{eqnarray}
The best-fit values of the asymmetries corresponding to $\chi^2_{min}$, are
presented in \cite{giun}:
\begin{eqnarray}
\label{eq:MM4}
A^{CPT}_{sin^2 2\theta}=0.42,~~A^{CPT}_{\delta m^2}=0.37 eV^2.  
\end{eqnarray}
Authors of  \cite{giun} consider that there is indication of a CPT-violating
asymmetry in experiments on a survival of the neutrino and the antineutrino from
the contradiction of the data from radioactive sources and reactor sources.

In our article \cite{efro} arguments against conclusions have been stated
article \cite{giun}.
In particular the attention has been paid that the technique of
\cite{malt} does not allow to estimate uncertainty of defined parametres.
Which can be rather more than uncertainty of
$\chi^2_{min}$ definition
\cite{coll}.

In the present article we add arguments about it and we define in the order of
size a dispersion 
of $\delta m^2_{\nu}$ and $\delta m^2_{\bar{\nu}}$.

First of all we notice from classification of neutrino oscillation experiments
by size of the relation
$L/E$ sensitivity of $\delta m^2$:

1. Experiments with short-baseline. In these experiments
$L/E \leq 1 eV^{-2}$, sensitivity $\delta m^2 \geq 0.1 eV^2$. Experiments
concern this type with radioactive sources and reactor experiments.

2. Experiments with long-baseline and atmospheric experiments. For them
$L/E \leq 10^4 eV^{-2}$, sensitivity $\delta m^2 \geq 10^{-4} eV^2$.
Examples of such experiments are reactor experiment CHOOZ, accelerator
experiment MINOS.

3. Experiments with very big flying base and solar experiments.
The example is reactor experiment
KamLand with $L \simeq 180 km$, $E \simeq 3 MeV$, thus
$L/E \simeq 3\cdot10^5 eV^{-2}$, and sensitivity
$\delta m^2 \geq 3\cdot10^{-5} eV^2$. For solar neutrino experiments 
GALLEX, GNO, SAGE, $L \simeq 1.5\cdot10^8 km$, $E \simeq 1 MeV$,
$L/E \sim 10^{12} eV^{-2}$, and sensitivity $\delta m^2 \geq 10^{-12} eV^2$. 

Procedure of calculation of oscillations parametres is difficult.
Nevertheless there is a question.

 Whether the same physical combination of neutrino masses can change at increase
 in flying base?
If can, then it any more that it is accepted to name masses.
Then it is not so elementary particles, and certain phantoms.

In a formula (\ref{eq:MM1}) conclusion it is supposed that all neutrinos in a
bunch possess the same fixed momentum with the big accuracy of $\delta m^2$
assuming definition. In experiments at long baseline the bunch is built in a
direction and on momentum size. Character of a bunch becomes more suitable to the
assumption of an applied formalism. From here and more precisely there is
$\delta m^2$ calculation. That has  no place at short baseline experiments.   

In the S-matrix theory the final condition is considered removed enough from a
reaction place. Then the right answer about $\delta m^2$ it is necessary to look
at least in experiments with long-baseline.
Further we consider that $\delta m^2$ in these experiments essentially is less,
than in experiments with short-baseline and we assume normal distribution of
$\delta m^2$.Then with probability
68,3\%, root from dispersion of $\delta m^2$ in experiments with 
short-baseline in the order of size will be equal to the parametre $\delta m^2$.

Further we will use the date of the review \cite{gonz} concerning
$\delta m^2_{min}$ from experiments with short-baseline for the channel
$\nu_{\mu} \rightarrow \nu_e$:
\begin{eqnarray}
\label{eq:mm5}
channel&&\delta m^2_{min}(eV^2)[Ref]\nonumber\\       
\nu_{\mu} \rightarrow \nu_e&&0.075 \cite{boro}\nonumber\\
\nu_{\mu} \rightarrow \nu_e&&0.4 \cite{ahre}\nonumber\\
\nu_{\mu} \rightarrow \nu_e&&1.6 \cite{romo,avva}\nonumber\\
\nu_{\mu} \rightarrow \nu_e&&0.4 \cite{asti}
\end{eqnarray}

We calculate root-mean-square value of ${\delta m^2_{\nu}}_{min}$ for four
experiments of the channel $\nu_{\mu} \rightarrow \nu_e$. Also we receive value 
for uncertainty $\delta m^2_{\nu}$ - $\sigma_{\nu}=0.42 eV^2$. Value for
uncertainty $\delta m^2_{\bar{\nu}}$ it is received from Ref. \cite{giunt},
where the interval for $\delta m^2_{\bar{\nu}}$ is
$0.2\leq\delta m^2_{\bar{\nu}}\leq2eV^2$.Then for uncertainty of
$\delta m^2_{\bar{\nu}}$ - we receive value -  $\sigma_{\bar{\nu}}=0.2eV^2$.
Uncertainty of asymmetry $A^{CPT}_{\delta m^2} (\ref{eq:MM3})~ \sigma_A$
is equal  
\begin{eqnarray}
\label{eq:MM6}
\sigma_A=\sqrt{0.42^2+0.2^2}=0.46 eV^2.  
\end{eqnarray}

Taking into account the equations (\ref{eq:MM4}) there is in our approach no
hint on CPT-violation from results of Ref. \cite{giun}.

In summary we do a conclusion that neutrino oscillation experiments with
short-baseline are not approaching at least for definition of CPT-violation.

\end{document}